\documentclass[epj]{svjour}
\usepackage{graphicx}
\usepackage{dcolumn}
\usepackage{bm}
\usepackage{setspace}
\usepackage{amssymb}
\usepackage{amsmath}
\usepackage[dvips]{color}
\usepackage{color}

\newcommand{\dblone}{\hbox{$1\hskip -1.2pt\vrule depth 0pt height 1.6ex width 0.7pt%
\vrule depth 0pt height 0.3pt width 0.12em$}}

\begin{document}

\title{Neutrino emissivity in the quark-hadron mixed phase of neutron stars}


\author{William M.\ Spinella\inst{1,2}, Fridolin Weber\inst{2,3}, Gustavo A.\
  Contrera\inst{4,5,6}, \and Milva G.\ Orsaria\inst{4,6}}

%
%
\institute{Computational Science Research Center San Diego State
  University, 5500 Campanile Drive, San Diego, CA 92182, USA.
\and 
  Department of Physics, San Diego State University, 5500 Campanile
  Drive, San Diego, CA 92182, USA.
\and
  Center for Astrophysics and Space Sciences, University of California
  San Diego, 9500 Gilman Drive, La Jolla, CA 92093, USA.
\and
CONICET, Rivadavia 1917, 1033 Buenos Aires, Argentina.
\and
IFLP, CONICET - Dpto. de F{\'i}sica, UNLP, La Plata, Argentina.
\and
Grupo de Gravitaci\'on, Astrof\'isica y Cosmolog\'ia, Facultad de
Ciencias Astron{\'o}micas y Geof{\'i}sicas, Universidad Nacional de La
Plata, Paseo del Bosque S/N (1900), La Plata, Argentina.}

\date{Received: date / Revised version: date}
%
\abstract{Numerous theoretical studies using various equation of state
  models have shown that quark matter may exist at the extreme
  densities in the cores of high-mass neutron stars. It has also been
  shown that a phase transition from hadronic matter to quark matter
  would result in an extended mixed phase region that would segregate
  phases by net charge to minimize the total energy of the phase,
  leading to the formation of a crystalline lattice. The existence of
  quark matter in the core of a neutron star may have significant
  consequences for its thermal evolution, which for thousands of years
  is facilitated primarily by neutrino emission.  In this work we
  investigate the effect a crystalline quark-hadron mixed phase can
  have on the neutrino emissivity from the core.  To this end we
  calculate the equation of state using the relativistic mean-field
  approximation to model hadronic matter and a nonlocal extension of
  the three-flavor Nambu-Jona-Lasinio model for quark matter. Next we
  determine the extent of the quark-hadron mixed phase and its
  crystalline structure using the Glendenning construction, allowing
  for the formation of spherical blob, rod, and slab rare phase
  geometries. Finally we calculate the neutrino emissivity due to
  electron-lattice interactions utilizing the formalism developed for
  the analogous process in neutron star crusts.  We find that the
  contribution to the neutrino emissivity due to the presence of a
  crystalline quark-hadron mixed phase is substantial compared to
  other mechanisms at fairly low temperatures ($\lesssim 10^9$ K) and
  quark fractions ($\lesssim 30\%$), and that contributions due to
  lattice vibrations are insignificant compared to static-lattice
  contributions.
} 
\authorrunning{{W. M. Spinella {\it et al.}}}
\titlerunning{Neutrino emissivity in quark-hadron mixed phase of
  neutron stars}
\maketitle
%

\maketitle

\section{\label{sec:intro}Introduction}
Seconds after the formation of a neutron star the mean free path of
neutrinos grows beyond the star's radius, and neutrinos from the core
escape easily cooling the new star rapidly. Neutrino emission
continues to be the dominant energy loss mechanism of a neutron star
for thousands of years until a temperature of about $10^7$ K is
reached \cite{NS1}.  It has been shown that the presence of quark
matter in the core of a neutron star can have a significant impact on
the neutrino emissivity, and suggested that this impact could have an
observable effect on the star's thermal evolution
\cite{weberPPNP,glendenning2001}.  In this work we investigate the
effect that a mixed phase containing quark matter may have on the
neutrino emissivity of a neutron star.

If electric charge neutrality in a neutron star is to be treated
globally as proposed by Glendenning \cite{glendenning1992}, then the
first order phase transition from hadronic matter to quark matter in
the core will result in a mixed phase in which both phases of matter
coexist. To minimize the total isospin asymmetry energy the two phases
will segregate themselves, resulting in positively charged regions of
hadronic matter and negatively charged regions of quark matter, with
the rare phase occupying sites on a Coulomb lattice. Further, the
competition between the Coulomb and surface energy densities will
cause the matter to arrange itself into energy minimizing geometric
configurations \cite{glendenning2001}.

The presence of the Coulomb lattice and the nature of the geometric
configurations of matter in the quark-hadron mixed phase may have a
significant effect on the neutrino emissivity from the core. More 
specifically, the presence of electrons in the mixed phase will lead to
an additional neutrino emissivity mechanism due to interactions with the
lattice. This process is analogous to neutrino-pair bremsstrahlung of
electrons in the neutron star crust, where ions exist on a lattice
immersed in an electron gas, and for which there exists a large body of
work (see, for example \cite{flowers,itoh84a,itoh84b,itoh84c,itoh84d,%
pethick1997,kaminker1999}). The situation is somewhat more complicated
in the quark-hadron mixed phase, but the operative interaction is
still the Coulomb interaction. Thus, to estimate the neutrino-pair 
bremsstrahlung of electrons in the quark-hadron mixed phase we
rely heavily on this body of work (particularly \cite{kaminker1999}).


Neutrino emissivity due to the interaction of electrons with a
crystalline quark-hadron mixed phase has been previously studied in
this manner by Na {\it et al.}\ \cite{Na}. In the present work we
replace the MIT Bag Model used by \cite{Na} to describe quark matter
with a three-flavor nonlocal variant of the Nambu-Jona-Lasinio model.
Next, we extend the range of possible geometric structures in the
mixed phase beyond spherical blobs to include rods and slabs, and
calculate the associated static lattice contributions to the neutrino
emissivity. Phonon contributions to the emissivity for rod and slab
geometries are not considered, though a comparison of the phonon and
static lattice contributions for spherical blobs is given and
indicates that phonon contributions may not be significant. Finally,
the extent of the conversion to quark matter in the core is determined
for the chosen parameterizations, and this allows for a targeted
comparison between emissivity contributions from standard neutrino
emission mechanisms (modified Urca, nucleon-nucleon and quark-quark
(NN+QQ) bremsstrahlung) and contributions from electron-lattice
interactions. In this work the minimal cooling paradigm is assumed,
as the mechanism under investigation is not expected to compete with
the direct Urca process, but may serve to enhance the cooling
of neutron stars in its absence.

This paper is structured as follows. In Sec. \ref{sec:eos} we discuss
the zero temperature equation of state of a neutron star 
containing hadronic matter, quark matter, and a quark-hadron mixed phase.
The crystalline structure of the mixed phase is described in Sec.
\ref{sec:structure}.  The neutrino emissivity due to interactions
between electrons and the crystalline lattice in the quark-hadron
mixed phase is described in Sec. \ref{sec:emissivity}. Our results,
including neutron star properties and neutrino emissivity
calculations, are presented in Sec. \ref{sec:results}.  Finally, we
present our conclusions in Sec.  \ref{sec:conclusions}.


\section{\label{sec:eos}Equation of State}


\subsection{Neutron Star Crust}
The neutron star outer and inner nuclear crust exists at densities
between $10^4$ g cm$^{-3} \lesssim \epsilon_{\mathrm{Crust}} \lesssim
10^{14}$ g cm$^{-3}$ \cite{weberbook}. Matter in the inner crust
consists mostly of nuclei in a Coulomb lattice that is immersed in a
gas of electrons and, above neutron drip ($\gtrsim 4\times10^{11}$ g
cm$^{-3}$), free neutrons.  In this work we use a combination of the
Baym-Pethick-Sutherland and Baym-Bethe-Pethick equations of state for
the nuclear crust \cite{BPS,BBP}. Compared to the neutron star core
the crust has less affect on the neutron star properties that are to
be studied in this work.


\subsection{Confined Hadronic Phase}

The hadronic phase of neutron star matter exists at densities above that of
the crust and is populated by baryons
($B=\{n,p,\Lambda,\Sigma,\Xi\}$) and leptons
($\lambda=\{e^-,\mu^-\}$).  To model the hadronic phase we use the
relativistic mean-field approximation (RMF), in which the interactions
between baryons are described by the exchange of scalar ($\sigma$),
vector ($\omega$), and isovector ($\rho$) mesons \cite{walecka}.  The
mean-field Lagrangian is given by
\cite{glen85:b,weberbook,glendenningbook,selfinteractions,boguta77:a,boguta83:a}
\begin{equation} \label{eq:lagrangian}
  \begin{aligned}
    \mathcal{L} = & \sum\limits_{B}
    \bar\psi_B\biggl[\gamma_{\mu}\left(i\partial^{\mu}
    -g_{\omega B}\omega^{\mu}-\frac{1}{2}g_{\rho B}\boldsymbol{\tau}\cdot
    \boldsymbol{\rho}^{\mu}\right)\\
    &-\left(m_n-g_{\sigma B}\sigma\right)\biggl]\psi_B
    +\frac{1}{2}(\partial_{\mu}\sigma\partial^{\mu}\sigma-
    m^2_{\sigma}\sigma^2)\\
    &-\frac{1}{3}b_{\sigma}m_n
    \left(g_{\sigma}\sigma\right)^3 - \frac{1}{4}c_{\sigma}
    \left(g_{\sigma}\sigma\right)^4 \\
    &- \frac{1}{4}\omega_{\mu\nu}
    \omega^{\mu\nu} 
    +\frac{1}{2}m^2_{\omega}\omega_{\mu}\omega^{\mu} \\
    &+\frac{1}{2}
    m^2_{\rho}\boldsymbol{\rho}_{\mu}\cdot
    \boldsymbol{\rho}^{\mu}
    -\frac{1}{4}\boldsymbol{\rho}_{\mu\nu}\cdot
    \boldsymbol{\rho}^{\mu\nu}\\
    &+\sum\limits_{\lambda}\bar\psi_{\lambda}\left(i\gamma_{\mu}
    \partial^{\mu}-m_\lambda\right)\psi_{\lambda} \, .
  \end{aligned}
\end{equation}
The $\sigma$ and $\omega$ mesons are responsible for nuclear binding
while the $\rho$ meson is required to obtain the correct value for the
empirical symmetry energy. In contrast to $\sigma$ and $\omega$
mesons, which are isoscalars, the $\rho$ meson is  an isovector
field  that manifests itself in the occurrence of the Pauli
matrix $\boldsymbol\tau$ ($ = (\tau^1,\tau^2,\tau^3)$) in
Eq.\ (\ref{eq:lagrangian}).  The cubic and quartic $\sigma$ terms in
Eq.\ (\ref{eq:lagrangian}) are necessary (at the relativistic
mean-field level) to obtain the empirical incompressibility of nuclear
matter \cite{selfinteractions,boguta77:a}. The field tensors
$\omega_{\mu\nu}$ and ${\boldsymbol\rho}_{\mu\nu}$ are defined as
$\omega_{\mu\nu} = \partial_\mu\omega_\nu - \partial_\nu\omega_\mu$
and ${\boldsymbol\rho}_{\mu\nu} = \partial_\mu{\boldsymbol\rho}_\nu -
\partial_\nu{\boldsymbol\rho}_\mu$.

The meson-baryon coupling constants ($g_{\sigma N}$, $g_{\omega N}$,
$g_{\rho N}$, $b_{\sigma}$, $c_{\sigma}$) of the Lagrangian are set so
that the properties of nuclear matter at saturation density are
reproduced for the appropriate parameterization (Table
\ref{table:parameterization}).
In this work we employ the GM1 and NL3
parameterizations as in Ref.\ \cite{NJL2014}.
To fix the meson-hyperon coupling
constants $g_{mY}$ we follow the method presented in Ref.
\cite{Miyatsu2013}.
The scalar meson-hyperon coupling constants $g_{\sigma Y}$ are fit to
the following hypernuclear potentials at saturation density:
$U_{\Lambda}^{(N)} = -28$ MeV, $U_{\Sigma}^{(N)} = +30$ MeV,
and $U_{\Xi}^{(N)} = -18$ MeV.
The vector meson-hyperon coupling constants $g_{\omega Y}$ are fixed
in SU(3) flavor symmetry by the mixing angle $\theta_v$ and coupling ratio
$z$ taken from the Nijmegen extended-soft-core (ESC08) model \cite{ESC08}.
The isovector meson-hyperon coupling constants $g_{\rho Y}$ are given
by the usual relations, $g_{\rho \Lambda} = 0$ and $g_{\rho \Sigma} =
2g_{\rho \Xi} = 2g_{\rho N}$.
\begin{table}
    \caption{Properties of nuclear matter at saturation density for
      the NL3 and GM1 parameterizations. Properties include the
      nuclear saturation density $\rho_0$, energy per nucleon $E/N$,
      compression modulus $K$, effective nucleon mass $m_N^*$, and
      asymmetry energy $a_{sy}$.}
    \label{table:parameterization}
    \begin{tabular}{ccc}
    \hline\noalign{\smallskip}
      Nuclear saturation properties & $~~$GM1$~~$ & $~~$NL3$~~$ \\
    \noalign{\smallskip}\hline\noalign{\smallskip}
      $\rho_0$ (fm$^{-3}$) & 0.153 & 0.148 \\
      $E/N$ (MeV) & -16.3 & -16.3\\
      $K$ (MeV) & 300 & 272 \\
      $m^*/m_N$ & 0.78 & 0.60 \\
      $a_{sy}$ (MeV) & 32.5 & 37.4 \\
   \noalign{\smallskip}\hline
   \end{tabular}
\end{table}

The field equations for the baryon fields follow from
Eq.\ (\ref{eq:lagrangian}) as follows
\cite{weberPPNP,glen85:b,weberbook,glendenningbook},
\begin{eqnarray}
\left( i \gamma^\mu\partial_\mu-m_B \right) \psi_B &=& - g_{\sigma B}
\sigma \psi_B + g_{\omega B}\gamma^\mu\omega_\mu \psi_B \nonumber
\\ &&+ g_{\rho B} \gamma^\mu
\ {\boldsymbol\tau}\boldsymbol\cdot{\boldsymbol\rho}_\mu \psi_B \, .
\label{eq:eompsi}
\end{eqnarray}
The meson fields in (\ref{eq:eompsi}) are solutions of the following
field equations \cite{weberPPNP,glen85:b,weberbook,glendenningbook},
\begin{eqnarray}
\bigl( \partial^\mu\partial_\mu+m^2_\sigma) \sigma &=& \sum_B
g_{\sigma B} \bar\psi_B \psi_B - m_N b_N g_{\sigma N} \left( g_{\sigma
  N} \sigma \right)^2 \nonumber \\ &&- c_N\, g_{\sigma N} \left(
g_{\sigma N} \sigma \right)^3 \, ,
\label{eq:eomsigma} \\
\partial^\mu \omega_{\mu\nu} + m_\omega^2 \, \omega_\nu &=& \sum_B
g_{\omega B} \bar\psi_B \gamma_\nu \psi_B \, ,
\label{eq:eomom5} \\
\partial^\mu {\boldsymbol\rho}_{\mu\nu} + m_\rho^2 \,
\boldsymbol\rho_\nu &=& \sum_B  g_{\rho B} \bar\psi_B
\boldsymbol\tau \gamma_\nu \psi_B \, .
\label{eq:eomrho5}
\end{eqnarray}
In the mean-field limit, the meson field equations (\ref{eq:eomsigma})
through (\ref{eq:eomrho5}) are given by
\cite{weberPPNP,glen85:b,weberbook,glendenningbook}
\begin{equation} \label{eq:sigma}
  \begin{aligned}
    m^2_{\sigma}\sigma = & \sum_B g_{\sigma B}\frac{2J_B+1}{2\pi^2}
    \int_0^{k_B}
    \frac{m_B^*(\sigma)}{\sqrt{k^2+m_B^{*2}(\sigma)}}\,k^2
    dk\\ &-bm_ng_{\sigma}(g_{\sigma}\sigma)^2-cg_{\sigma}(g_{\sigma}\sigma)^3 \, ,
    \end{aligned}
\end{equation}
\begin{equation} \label{eq:omega}
  \omega_0 = \sum_B \frac{g_{\omega B}}{m_{\omega}^2}\rho_B \, ,
\end{equation}
\begin{equation} \label{eq:rho}
  \rho_{03} = \sum_B \frac{g_{\rho B}}{m_{\rho}^2}I_{3B}\rho_B \, ,
\end{equation}
where the effective baryon mass $m^*_B(\sigma) = m_B - g_{\sigma B}\sigma$.

To determine the equation of state we solve a nonlinear system
consisting of the meson mean-field equations and the charge
conservation conditions (baryonic, electric) given by
\cite{glen85:b,weberbook,glendenningbook}
\begin{equation} \label{eq:baryonic}
  \rho_b-\sum\limits_B\rho_B=0 \, ,
\end{equation}
\begin{equation} \label{eq:charge}
  \sum\limits_{B}\rho_Bq_B
  +\sum\limits_{\lambda}\rho_{\lambda}q_{\lambda}=0 \, ,
\end{equation}
where $\rho_b$ is the total baryonic density and $q_B$ and $q_\lambda$
are the electric charges of baryons and leptons, respectively.
Particles in the hadronic phase are subject to the chemical
equilibrium condition,
\begin{equation} \label{eq:chemicalequilibrium}
  \mu_i = b_i\mu_N - q_i\mu_e \, ,
\end{equation}
where $\mu_i$ is the chemical potential and $b_i$ is the baryon number
of particle $i$.  New baryon or lepton states are populated when the
right side of equation (\ref{eq:chemicalequilibrium}) is greater than
the states' chemical potential. The baryonic and leptonic number
densities ($\rho_B$, $\rho_{\lambda}$) are both given by
\begin{equation} \label{eq:numberdensity}
  \rho_i = (2J_i+1)\frac{k_i^3}{6\pi^2} \, .
\end{equation} 
The free parameters of the system are the meson mean-fields ($\sigma$,
$\omega$, $\rho$), and the neutron and electron fermi momenta ($k_n$,
$k_e$). Finally, the energy density and pressure of the hadronic phase
are given by \cite{glen85:b,weberbook,glendenningbook}
\begin{equation} \label{eq:hdensity}
  \begin{aligned} 
    \epsilon_H = &\,\frac{1}{3}bm_n(g_{\sigma}\sigma)^3
    +\frac{1}{4}c(g_{\sigma}\sigma)^4
    +\frac{1}{2}m_{\sigma}^2\sigma^2 \\
    & +\frac{1}{2}m_{\omega}^2\omega_0^2
    +\frac{1}{2}m_{\rho}^2\rho_{03}^2\\
    &+\sum_B \frac{2J_B+1}{2\pi^2}
    \int_0^{k_B}\sqrt{k^2+m_B^{*2}(\sigma)}\,k^2 dk\\
    &+\sum_{\lambda} \frac{1}{\pi^2} \int_0^{k_{\lambda}}
    \sqrt{k^2+m_{\lambda}^2}\,k^2 dk \, ,
  \end{aligned}
\end{equation}
\begin{equation} \label{eq:hpressure}
  \begin{aligned}
    p_H = &\,-\frac{1}{3}bm_n(g_{\sigma}\sigma)^3
    -\frac{1}{4}c(g_{\sigma}\sigma)^4
    -\frac{1}{2}m_{\sigma}^2\sigma^2 \\
   & +\frac{1}{2}m_{\omega}^2\omega_0^2
    +\frac{1}{2}m_{\rho}^2\rho_{03}^2\\
    &+\frac{1}{3}\sum_B \frac{2J_B+1}{2\pi^2}
    \int_0^{k_B}\frac{k^4 dk}{\sqrt{k^2+m_B^{*2}(\sigma)}}\\
    &+\frac{1}{3}\sum_{\lambda} \frac{1}{\pi^2} \int_0^{k_{\lambda}}
    \frac{k^4 dk}{\sqrt{k^2+m_{\lambda}^2}} \, .
  \end{aligned}
\end{equation}


\subsection{Deconfined Quark Phase}

If the dense interior of a neutron star contains deconfined quark
matter, it will be made of up ($u$), down ($d$), and strange ($s$)
quarks in chemical equilibrium with a small number of electrons and
muons.  To model the quark phase we use a nonlocal extension of the
Nambu-Jona-Lasinio model (n3NJL) as described in Ref.\ \cite{NJL2014}.
The effective action of this model is given by
\begin{eqnarray}
S_E &=& \int d^4x \Bigl\{\bar \psi (x)( i \partial{\hskip-2.0mm}/ -
{\hat m}) \psi (x) \;+\; \frac{1}{2}\,\,G_S [\,({\bar \psi
    (x)}\lambda_a \psi (x))^2 \nonumber \\
  &&+ ({\bar \psi (x)} i\gamma_5\lambda_a
  \psi (x))^2\,] + H \,\bigl[ \,{\rm det} [{\bar \psi
      (x)}(1+\gamma_5) \psi (x)] \nonumber \\
 && + {\rm det} [{\bar \psi
      (x)}(1-\gamma_5) \psi (x)] \,\bigr]  - G_V
[({\bar \psi (x)}\gamma^\mu\lambda_a \psi (x))^2 \nonumber \\
&&+ ({\bar \psi (x)}
  i\gamma^\mu \gamma_5\lambda_a \psi (x))^2]\,\Bigr\} ,
 \label{L3}
\end{eqnarray}
where $f$ denotes quark flavor ($u,d,s$), $\psi$ is a chiral U(3)
vector that includes the light quark fields, $\psi \equiv (u, d,
s)^T$, $\hat m = {\rm diag}(m_u, m_d, m_s)$ is the current quark mass
matrix, $\lambda_a$ with $a=1,...,8$ denote the generators of SU(3),
and $\lambda_0=\sqrt{2/3}\, \dblone_{3\times 3}$. The coupling constants
$G_S$ and $H$, the strange quark mass
$m_{s}$, and the three-momentum ultraviolet cutoff parameter
$\Lambda$, are all model parameters. Their values are taken from
Ref.\ \cite{rehberg}, i.e., $m_u=m_d=5.5$ MeV, $m_s=140.7$ MeV,
$\Lambda=602.3$ MeV, $G_S\Lambda^2=3.67$ and $H\Lambda^5=-12.36$. The
vector coupling constant $G_V$ is treated as a free parameter.

For the mean-field approximation, the thermodynamic
potential associated with $S_E$ of Eq.\ (\ref{L3}) is given by
\begin{equation} \label{grandpotential}
  \begin{aligned}
    &\Omega^{\mathrm{NL}} = -\frac{3}{\pi^3}\sum_{f=u,d,s}
    \int_0^{\infty}dp_0 \int_0^{\infty} dp\\ &~~~~~~~~~~~~~~~~~~~~~~~~~~~\times \mathrm{ln}
    \Biggl\{ \left[ \widehat{\omega}^2_f+M_f^2(\omega^2_f)\right]
    \frac{1}{\omega^2_f+m^2_f}\Biggr\} \\
    &\quad -\frac{3}{\pi^2}\sum_{f=u,d,s}\int_0^{\sqrt{\mu^2_f-m^2_f}}
    dp\,p^2\left[\left(\mu_f-E_f\right)\theta\left(\mu_f-m_f\right)\right]\\
    &\quad -\frac{1}{2}\Bigg[\sum_{f=u,d,s}\left(\bar{\sigma}_f\bar{S}_f+
    \frac{G_S}{2}\bar{S}^2_f\right)+\frac{H}{2}\bar{S}_u\;\bar{S}_d\;\bar{S}_s
    \Biggr] \\
    &\quad - \sum_{f=u,d,s}\frac{\overline{\omega}_f^2}{4G_V} \, ,
  \end{aligned}
\end{equation}
where $\bar{\sigma}_f$, $\overline{\omega}_f$, and $\bar{S}_f$ are the
quark scalar, vector, and auxiliary mean fields, respectively.
Moreover, we have $E_f= \sqrt{{\boldmath p}^2 + m^2_f}$, $\omega^2_f =
(p_0+i\mu_f)^2 + {\boldsymbol p}^2$, and $M_f(\omega_f^2) =
m_f+\bar{\sigma}_f g(\omega^2_f)$ are the momentum dependent quark
masses. The quantity $g(\omega^2_f) =
\mathrm{exp}(-\omega^2_f/\Lambda^2)$ is the form factor which
introduces nonlocality into the quark interactions \cite{NJL2014}.
The auxiliary mean fields are given by
\begin{equation}
  \bar{S}_f = -48 \int_0^{\infty} dp_0 \int_0^{\infty}
  \frac{dp}{8\pi^3}g(\omega^2_f)\frac{M_f(\omega^2_f)}
  {\widehat{\omega}^2+M_f^2(\omega^2_f)} \, .
\end{equation}
Due to the inclusion of the vector interaction the quark chemical
potentials are shifted as follows,
\begin{equation} \label{eq:shiftedmu}
  \widehat{\mu}_f = \mu_f-g(w_f^2)\overline{\omega}_f \, ,
\end{equation}
\begin{equation} \label{eq:shiftedomega}
  \widehat{\omega}_f^2 = (p_0 + i \widehat{\mu}_f)^2+p^2 \, .
\end{equation}
The scalar and vector mean fields are obtained by minimizing the
grand thermodynamic potential,
\begin{equation}
  \frac{\partial \Omega^{\mathrm{NL}}}{\partial \bar{\sigma}_f} = 0 \, ,
  ~~~~~
  \frac{\partial \Omega^{\mathrm{NL}}}{\partial \overline{\omega}_f} = 0 \, .
\end{equation}
The quark number densities are given by
\begin{equation} \label{eq:qnumberdnesity}
  \rho_f = \frac{\partial \Omega^{\mathrm{NL}}}{\partial \mu_f} \, .
\end{equation}

To determine the equation of state one must solve a nonlinear system
of equations for the fields $\bar{\sigma}_f$ and $\bar{\omega}_f$, and
the neutron and electron chemical potentials $\mu_n$ and $\mu_e$.
This system of equations consists of the mean field equations,
\begin{equation} \label{eq:q1}
   \bar{\sigma}_i + G_S \bar{S}_i + \frac{1}{2}H\bar{S}_j \bar{S}_k = 0 \, ,
\end{equation}
with cyclic permutations over the quark flavors, 
\begin{equation} \label{eq:q4}
  \overline{\omega}_f -2 G_V \frac{\partial
    \Omega^{\mathrm{NL}}}{\partial \omega_f} = 0 \, , 
\end{equation}
and the charge conservation equations,
\begin{equation} \label{eq:qbaryonic}
  \sum_{f=u,d,s} \rho_f - 3 \rho_b = 0 \, ,
\end{equation}
\begin{equation} \label{eq:qcharge}
  \sum_{f=u,d,s} \rho_fq_f + \sum_{\lambda=e^-,\mu^-}
  \rho_{\lambda}q_{\lambda} = 0 \, .
\end{equation}
Finally, the pressure and energy density are given by
\begin{equation} \label{eq:qpressure}
  p_Q = \Omega_0-\Omega^{\mathrm{NL}} \, ,
\end{equation}
\begin{equation} \label{eq:qdensity}
  \epsilon_Q = -p_Q +
  \sum_{f=u,d,s} \rho_f \mu_f + \sum_{\lambda=e^-,\mu^-} \rho_{\lambda} \mu_{\lambda} \, ,
\end{equation}
where $\Omega_0$ is the grand thermodynamic potential $\Omega^{NL}$ calculated
for $\mu_f = \overline{\omega}_f = 0$.


\subsection{Quark-Hadron Mixed Phase}

When the pressure in the hadronic phase grows to a level equal to that
of the quark phase at the same baryonic density a first order phase
transition from hadronic matter to quark matter may begin.  Since a
theory that can treat both the hadronic and quark phases
simultaneously is currently unavailable, we construct the mixed phase
by blending RMF and n3NJL. Each phase is solved for separately, and
then the two are blended together under the Gibbs condition, $p_H =
p_Q$.  The pressure ($p_M$) and energy density ($\epsilon_M$) in the
mixed phase are given by \cite{glendenning2001,glendenning1992}
\begin{equation}
  p_M = \frac{1}{2}(p_H + p_Q)\, ,
\end{equation}
and
\begin{equation}
  \epsilon_M = (1-\chi)\epsilon_H + \chi\epsilon_Q \, ,
\end{equation}
where $\chi = V_Q/V_{\mathrm{Total}}$ is the quark fraction of the
mixed phase. Other properties such as the particle number densities can be
handled in a similar fashion.
\begin{figure}
  \resizebox{0.48\textwidth}{!}{%
    \includegraphics{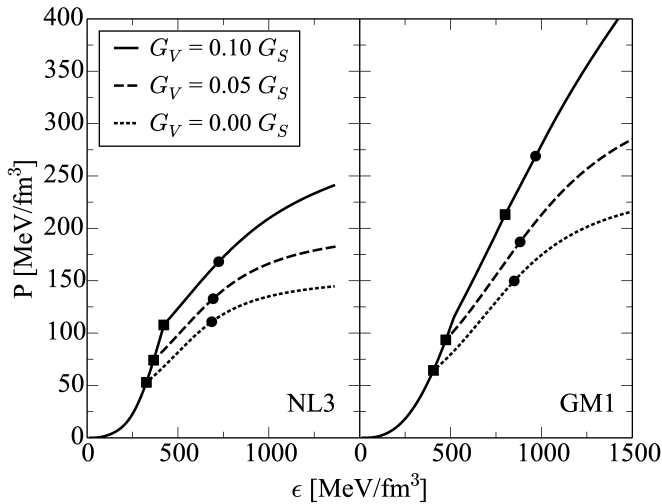}
    }
  \caption{Equations of state of this work. Shown are the hadronic 
	  and mixed phases. Square markers indicate the
	  beginning of the quark-hadron mixed phase, and dot markers indicate
	  the location of the maximum mass.}
  \label{fig:eos}
\end{figure}


\section{\label{sec:structure}Crystalline Structure of the Quark-Hadron Mixed Phase}

A mixed phase of hadronic and quark matter will arrange itself so as
to minimize the total energy of the phase. Under the condition of
global charge neutrality this is the same as minimizing the
contributions to the total energy due to phase segregation, which
includes the surface and Coulomb energy contributions.  Expressions
for the Coulomb ($\epsilon_C$) and surface ($\epsilon_S$) energy
densities can be written as \cite{glendenning2001}
\begin{eqnarray}
  \epsilon_C &=& 2\pi e^2 \left[ q_H(\chi) - q_Q(\chi)
    \right]^2 r^2 x f_D(x) \, ,
    \label{eq:eps_c} \\
    \epsilon_S &=& D x \sigma(\chi)/r\, ,
    \label{eq:eps_s}
\end{eqnarray}
where $q_H$ ($q_Q$) is the hadronic (quark) phase charge density, and
$r$ is the radius of the rare phase structure. The quantities $x$ and
$f_D(x)$ in Eq.\ (\ref{eq:eps_c}) are defined as
\begin{equation}
  x = \mathrm{min}(\chi,1-\chi) 
\end{equation}
and
\begin{equation}
  f_D(x) = \frac{1}{D+2}\left[\frac{1}{D-2}(2-Dx^{1-2/D})+x\right] \, ,
  \end{equation}
where $D$ is the dimensionality of the lattice. The quantity
$\sigma(\chi)$ in Eq.\ (\ref{eq:eps_s}) denotes the surface tension.

\begin{figure}[htb]
  \centering
  \includegraphics[width=0.95\linewidth]{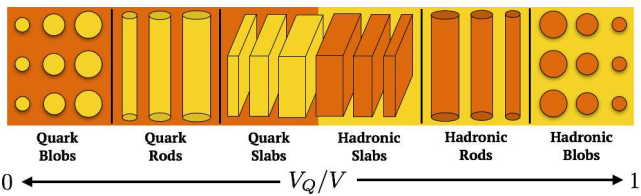}
  \caption{(Color online) Spherical blob, rod, and slab rare phase
    structures.  $V_Q/V_{\mathrm{Total}}$ denotes the
    quark fraction of the mixed phase.}
  \label{fig:shapes}
\end{figure}

The phase rearrangement process will result in the formation of
geometrical structures of the rare phase distributed in a crystalline
lattice that is immersed in the dominant phase (Figure
\ref{fig:shapes}).  The rare phase structures are approximated for
convenience as spherical blobs, rods, and slabs
\cite{glendenning2001}. The spherical blobs occupy sites in a three
dimensional ($D=3$) body centered cubic (BCC) lattice, the rods in a
two dimensional ($D=2$) triangular lattice, and the slabs in a simple
one dimensional ($D=1$) lattice \cite{kaminker1999}. At $\chi = 0.5$
both hadronic and quark matter exist as slabs in the same proportion,
and at $\chi > 0.5$ the hadronic phase becomes the rare phase with its
geometry evolving in reverse order (from slabs to rods to blobs).

\begin{figure*}[htb]
  \centering
  \includegraphics[width=0.7\linewidth]{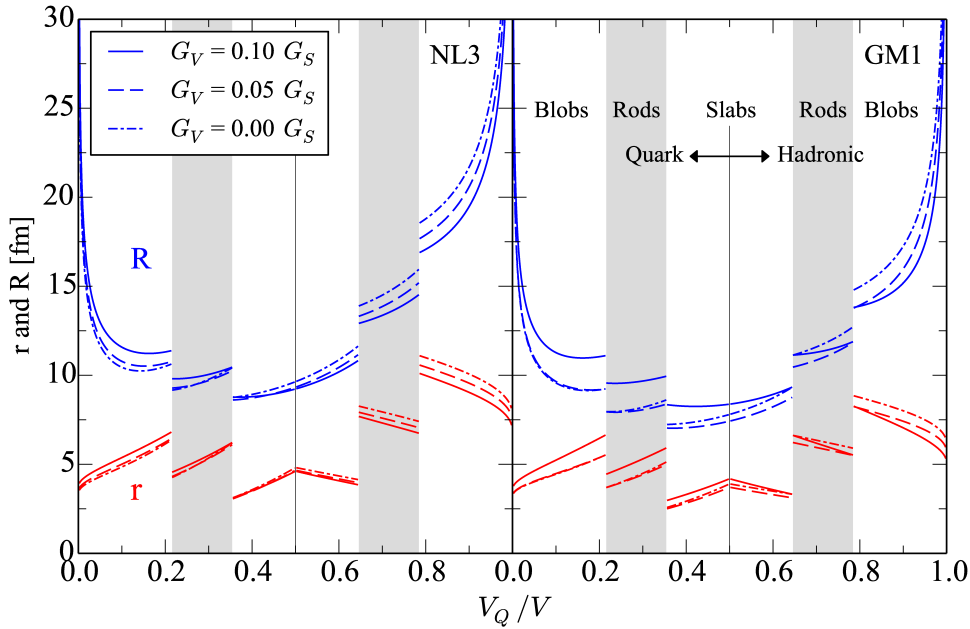}
  \caption{(Color online) Rare phase structure radius $r$ and Wigner-Seitz cell
    radius $R$ plotted against quark fraction for the given parameterizations
    and vector coupling constant values. The change in background color refers
  to a change in structure as shown in the right panel. Discontinuities
  in the radii are also associated with changes in the crystalline structure.}
  \label{fig:radii}
\end{figure*}

\subsection{Surface Tension of the Quark-Hadron Interface}

Direct determination of the surface tension of the quark-hadron interface 
is problematic because of difficulties in constructing a single theory that can 
accurately describe both hadronic matter and quark matter.
Therefore, we employ an approximation proposed by Gibbs where the
surface tension is taken to be proportional to the difference in the energy
densities of the interacting phases \cite{glendenning2001},
\begin{equation}
  \sigma(\chi)=\eta L \left[\epsilon_Q(\chi)-\epsilon_H(\chi)\right]\, ,
\end{equation}
where $L$ is proportional to the surface thickness which should be on the order of the
strong interaction (1 fm), and $\eta$ is a proportionality constant. In this work
we maintain the energy density proportionality but set the parameter $\eta$
so that the surface tension falls below 50 MeV fm$^{-2}$,
a value consistent with those suggested for $\sigma(\chi)$ in recent literature
\cite{yasutake2014,surfacetension1,surfacetension2,surfacetension3}. 

\subsection{Rare Phase Structure Size, Charge, and Number Density}

The size of the rare phase structures is given by the radius ($r$)
and is determined by minimizing the sum of the Coulomb and surface
energies, $\frac{\partial(\epsilon_C+\epsilon_S)}{\partial r}$, and solving
for $r$ \cite{glendenning2001}, 
\begin{equation}
  r = \left(\frac{D\sigma(\chi)}
  {4\pi e^2f_D(\chi)\left[q_H\left(\chi\right)
  - q_Q(\chi)\right]^2}\right)^\frac{1}{3}.
\end{equation}
The primitive cell of the lattice is taken to be the Wigner-Seitz cell,
though it is simplified to have the same geometry as the rare phase
structure. The Wigner-Seitz cell radius $R$ is set so that the cell is
charge neutral.

\begin{figure}
  \centering
  \includegraphics[width=0.95\linewidth]{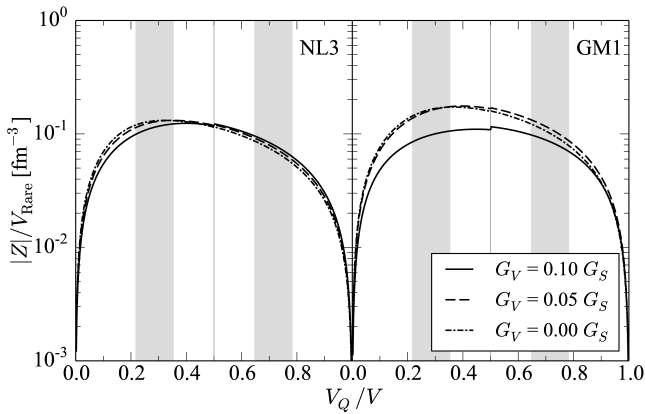}
  \caption{(Color online) Rare phase structure charge number per unit
    volume plotted against quark fraction for the given
    parameterizations and vector coupling constant values. Slight
    discontinuity at $\chi=0.5$ is due to a higher number of
    positively charged baryons (electrons make up difference in
    negative charge).}
  \label{fig:charge}
\end{figure}

The density of electrons in the mixed phase is taken to be uniform
throughout. Charge densities in both the rare and dominant phases
are also taken to be uniform, an approximation supported by a recent study
by Yasutake {\it et al.}\ \cite{yasutake2014}. The uniformity of charge
in the rare phase also justifies the use of the nuclear form factor
($F(q)$) presented in Section \ref{sec:emissivity}.
The total charge number per unit volume ($\left|Z\right|/V_{\mathrm{Rare}}$)
of the rare phase structures is given in Figure \ref{fig:charge}.

The number density of rare phase blobs will be important for calculating
the phonon contribution to the emissivity. Since there is one rare phase
blob per Wigner-Seitz cell, the number density of rare phase blobs ($n_b$)
is simply the reciprocal of the Wigner-Seitz cell volume,
\begin{equation} \label{eq:blobdensity}
  n_b = (4\pi R^3/3)^{-1} \, .
\end{equation}


\section{\label{sec:emissivity}Neutrino Emissivity in the Quark-Hadron Mixed Phase}

Modeling the complex interactions of electrons with a background of
neutrons, protons, hyperons, muons, and quarks is an exceptionally
complicated problem. However, to make a determination of the neutrino
emissivity that is due to electron-lattice interactions in the
quark-hadron mixed phase we need only consider the Coulomb interaction
between them.  This simplifies the problem greatly, as a significant
body of work exists for the analogous process of electron-ion
scattering that takes place in the crusts of neutron stars.



\subsection{Electron-Lattice Interaction}

To determine the state of the lattice in the quark-hadron mixed
phase we use the dimensionless ion coupling parameter given by \cite{NS1}
\begin{equation}
  \Gamma = \frac{Z^2e^2}{Rk_bT}\, .
\end{equation}
Below $\Gamma_{\mathrm{melt}} = 175$ the lattice behaves as a Coulomb
liquid, and above as a Coulomb crystal \cite{NS1}.  It was shown in Na
{\it et al.}\ \cite{Na} that the emissivity due to electron-blob interactions
in the mixed phase was insignificant compared to other contributions
\begin{figure}[htb]
  \centering
  \includegraphics[width=0.95\linewidth]{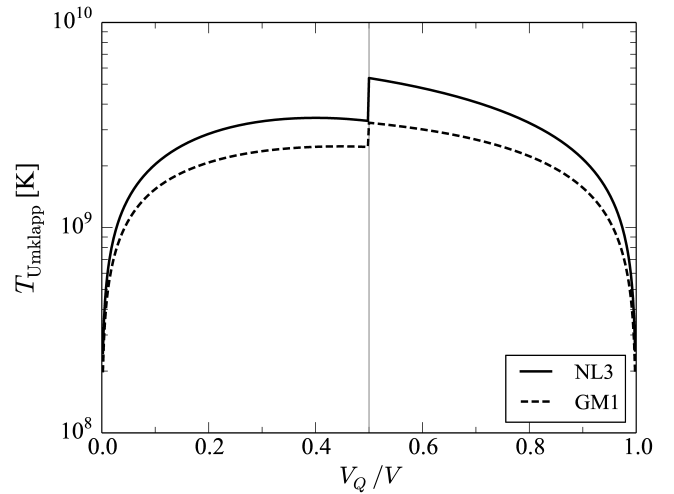}
  \caption{Temperature below which Umklapp processes are frozen out
	  ($T_{\mathrm{Umklapp}}$) as a function of quark fraction,
	  and contributions to the neutrino emissivity due to electron-phonon
  interactions become negligible.}
  \label{fig:umklapp}
\end{figure}
at temperatures above $T \gtrsim 10^{10}$ K. Therefore, in this work
we consider temperatures in the range $10^7\,\mathrm{K} \le T \le
10^{10}\,\mathrm{K}$. At these temperatures the value of the ion
coupling parameter is generally well above $\Gamma_{\mathrm{melt}}$,
and so the lattice in the quark-hadron mixed phase is taken to be a
Coulomb crystal.

To account for the fact that the elasticity of scattering events is
temperature dependent we need to compute the Debye-Waller factor,
which is known for spherical blobs only and
requires the plasma frequency and temperature given by
\begin{equation}
  \omega_p = \sqrt{\frac{4\pi Z^2e^2n_b}{m_b}}\, ,
\end{equation}
\begin{equation}
  T_p = \frac{\hbar \omega_p}{k_b}\, ,
\end{equation}
where $m_b$ is the mass of a spherical blob \cite{kaminker1999}.
The Debye-Waller factor is then given by 
\begin{equation} \label{eq:debye}
  W(q) = \begin{cases}
    \frac{\alpha q^2}{8k_e^2}\! \left(1.399\,e^{-9.1 t_p}+12.972\,t_p\right)
    & \mathrm{spherical~blobs}\, , \\
    0 & \mathrm{rods~and~slabs}\, ,
  \end{cases}
\end{equation}
where $q=|\boldsymbol{q}|$ is a phonon or scattering wave vector,
$\alpha = 4\hbar^2k_e^2/(k_B T_p m_b)$,
and $t_p = T/T_p$ \cite{kaminker1999,baiko1995}. In order to smooth out
the charge distribution over the radial extent of the 
rare phase structure we adopt the nuclear form factor given in \cite{kaminker1999},
\begin{equation} \label{eq:formfactor}
  F(q) =
  \frac{3}{(qR^3)}\left[\mathrm{sin}(qR)-qR\,\mathrm{cos}(qR)\right]
  \, .
\end{equation}
Screening of the Coulomb potential by electrons is taken into account
by the static dielectric factor $\epsilon(q,0)=\epsilon(q)$, given in
Ref.\ \cite{itoh1983_1}.  However, the charge number of the rare phase
structures is high and the electron number density is low, so setting
this factor to unity has no noticeable effect on the calculated
neutrino emissivity. Finally, the effective interaction is given by
\cite{kaminker1999}
\begin{equation} \label{eq:potential}
  V(q) = \frac{4\pi e \rho_Z F(q)}{q^2 \epsilon(q)}e^{-W(q)} \, .
\end{equation}


\subsection{Neutrino Emissivity}

General expressions for the neutrino emissivity due to electron-lattice
interactions were derived by Haensel {\it et al.}\ \cite{haensel1996} for spherical
blobs and by Pethick {\it et al.}\ \cite{pethick1997} for rods and slabs,
\begin{equation}
  Q_{\text{blobs}} \approx 3.35\times10^{-67}\,n_bT^6Z^2L
  \;\;\text{MeV\,s}^{-1}\text{\,fm}^{-3}\\[5pt]\, ,
\end{equation}
\begin{equation}
  Q_{\text{rods,slabs}} \approx 3.00\times10^{-88}\;k_eT^8J
  \;\;\text{MeV\,s}^{-1}\text{\,fm}^{-3}\, ,
\end{equation}
where $L$ and $J$ are dimensionless quantities that scale
the emissivities. Both $L$ and $J$ contain a contribution due
to the static lattice (Bragg scattering), but we consider the
additional contribution from lattice vibrations (phonons) for
spherical blobs, so $L = L_{\mathrm{sl}} + L_{\mathrm{ph}}$.


\subsection{Phonon Contribution to Neutrino Emissivity}

The expressions for determining the neutrino emissivity due to
interactions between electrons and lattice vibrations (phonons) in a
Coulomb crystal, with proper treatment of multi-phonon processes, were
obtained by Baiko {\it et al.}\ \cite{baiko1998} and simplified by Kaminker
{\it et al.}\ \cite{kaminker1999}. The phonon contribution to the
emissivity is primarily due to Umklapp processes in which a phonon is
created (or absorbed) by an electron that is simultaneously Bragg
reflected, resulting in a scattering vector $\boldsymbol{q}$ that lies
outside the first Brillouin zone, $q_0 \gtrsim (6\pi^2n_b)^{1/3}$
\cite{ziman,raikh1983}, where $n_b$ is given by
Eq.\ (\ref{eq:blobdensity}).

The contribution to the neutrino emissivity due to phonons is
contained in $L_{ph}$ and given by Eq.\ (21) in
Ref.\ \cite{kaminker1999},
\begin{equation} \label{eq:lph}
  L_{\mathrm{ph}} = \int_{y_0}^1 dy \frac{S_{\mathrm{eff}}(q)
  |F(q)|^2}{y|\epsilon(q,0)|^2}\left(1+\frac{2y^2}{1-y^2}
  \mathrm{ln}\,y\right)\, ,
\end{equation}
where $y=q/(2k_e)$, and the lower integration limit $y_0$ excludes
momentum transfers inside the first Brillouin zone.  The structure
factor $S_{\mathrm{eff}}$ is given by Eqs. (24) and (25) in
Ref.\ \cite{kaminker1999}),
\begin{equation} \label{eq:seff}
  \begin{aligned}
  S_{\mathrm{eff}}(q) = 189\left(\frac{2}{\pi}\right)^5 e^{-2W}
  \int_0^{\infty} d\xi \, &\frac{1-40\xi^2+80\xi^4}
  {\left(1+4\xi^2\right)^5
    \mathrm{cosh}^2\left(\pi\xi\right)} \\
&  \times \left(e^{\Phi(\xi)}-1\right)\, ,
  \end{aligned}
\end{equation}
\begin{equation} \label{eq:phi}
  \Phi(\xi) = \frac{\hbar q^2}{2m_b}
  \left\langle \frac{\mathrm{cos}\left(\omega_st\right)}
  {\omega_s \mathrm{sinh}\left(\hbar \omega_s/2k_BT\right)} \right\rangle\, ,
\end{equation}
where $\xi = tk_BT/\hbar$ and $\langle \ldots \rangle$ denotes
averaging over phonon frequencies and modes,
\begin{equation} \label{eq:fs}
  \langle f_s(\boldsymbol{k}) \rangle = \frac{1}{3V_B}\sum\limits_s
  \int_{V_B} d\boldsymbol{k}\, f_s(\boldsymbol{k})\, .
\end{equation}
It is assumed that there are three phonon modes $s$, two linear
transverse and one longitudinal. The frequencies of the transverse
\begin{figure}[htb]
  \centering
  \includegraphics[width=0.95\linewidth]{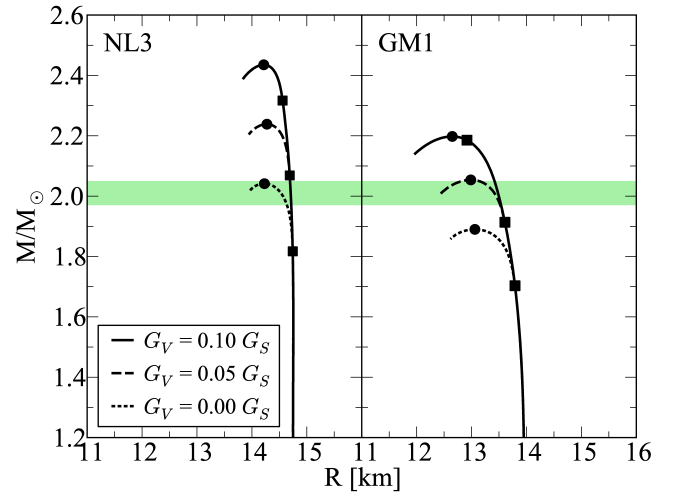}
  \caption{(Color online) Mass-radius relationship of neutron stars
    for the given parameterizations and vector coupling constant
    values listed in Table \ref{table:parameterization}. Square markers
    indicate the beginning of the quark-hadron mixed phase, and dot
    markers indicate the location of the maximum mass. The green
    shaded region indicates the mass constraint set by PSR J3048+0432.}
  \label{fig:tov}
\end{figure}
modes are given by $\omega_{t_i} = a_ik$, where $i=1,2$, $a_1 =
0.58273$, and $a_2 = 0.32296$.  The frequency of the longitudinal mode
$\omega_l$ is determined by Kohn's sum rule, $\omega_l^2 = \omega_p^2
- \omega_{t_1}^2 - \omega_{t_2}^2$ \cite{mochkovitch1979}.

Umklapp processes proceed as long as the temperature 
$T_{\mathrm{Umklapp}} \gtrsim T_pZ^{1/3}e^2/(\hbar c)$, below which electrons 
can no longer be treated in the free electron approximation \cite{raikh1983}.
This limits the phonon contribution to the neutrino emissivity
to only a very small range in temperature for a crystalline quark-hadron
mixed phase (see Figure \ref{fig:umklapp}), and renders it negligible compared to
the static lattice contribution as will be shown in the next section.


\subsection{Static Lattice Contribution to Neutrino Emissivity}

Pethick and Thorsson \cite{pethick1997} found that with proper
handling of electron band-structure effects the static lattice
contribution to the neutrino emissivity in a Coulomb crystal was
significantly reduced compared to calculations performed in the free
electron approximation. Kaminker {\it et al.}\ \cite{kaminker1999} presented
simplified expressions for calculating the static lattice contribution
($L_{\mathrm{sl}}$) using the formalism developed in
Ref.\ \cite{pethick1997}.  The dimensionless quantities
$L_{\mathrm{sl}}$ and $J$ that scale the neutrino emissivities for
spherical blobs and rods/slabs, respectively, are given by
\begin{equation} \label{eq:lsl}
  L_{\mathrm{sl}} = \frac{1}{12Z} \sum_{K \ne 0}
  \frac{(1-y_K^2)}{y_K^2}\frac{|F(K)|^2}{|\epsilon(K)|^2}\,I(y_K,t_V)
  \,e^{-2W(K)} 
\end{equation}
and
\begin{equation} \label{eq:j}
  J = \sum_{K \ne 0}\frac{y_K^2}{t^2_V}I(y_K,t_V)\, ,
\end{equation}
where $K=|\boldsymbol{K}|$ is a scattering vector and restricted to
linear combinations of reciprocal lattice vectors, $y_K=K/(2k_e)$,
$t_V = k_BT/\left[|V(K)|(1-y_K^2)\right]$, and $I(y_K,t_V)$ is given
by Eq.\ (39) in Ref.\ \cite{kaminker1999}.  The sum over $K$ in
Eqs. (\ref{eq:lsl}) and (\ref{eq:j}) terminates when $K > 2k_e$,
prohibiting scattering vectors that lie outside the electron Fermi
surface.


\section{\label{sec:results}Results}

The neutron star equation of state has been calculated using the
relativistic
mean field approximation to describe the hadronic phase and the
three-flavor nonlocal Nambu-Jona-Lasinio model for the quark 
phase, with the Gibbs condition governing the combination of the
two in the mixed phase (Figure \ref{fig:eos}).
Using the equation of state we solve the
Tolman-Oppenheimer-Volkoff equation \cite{tov,tov2} and find the mass-radius
relationships given in Figure \ref{fig:tov}. The
maximum masses of the neutron stars obtained
for the given parameter sets are able to account for the recently
discovered high mass pulsars PSR J3048+0432 and PSR J1614-2230
\cite{psrj1614,psrj3048,psrj3048b}, excluding GM1 with no vector coupling.
It is evident from Figure \ref{fig:tov} that increasing
the vector coupling constant increases the maximum mass for the
particular parameterization.
\begin{figure}[htb]
  \centering
  \includegraphics[width=0.95\linewidth]{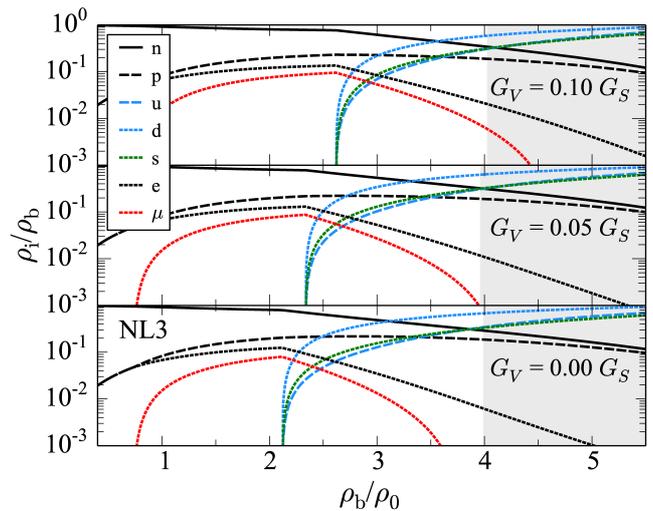}
  \caption{(Color online) Relative particle number densities as a
    function of baryonic density (in units of nuclear saturation
    density) for the NL3 parameterization and given values of the
    vector coupling constant. Grey shaded region indicates densities
    beyond the maximum mass neutron star for the given parameterization.}
  \label{fig:nl3_numberdensity} 
\end{figure}
\begin{figure}[htb]
  \centering
  \includegraphics[width=0.95\linewidth]{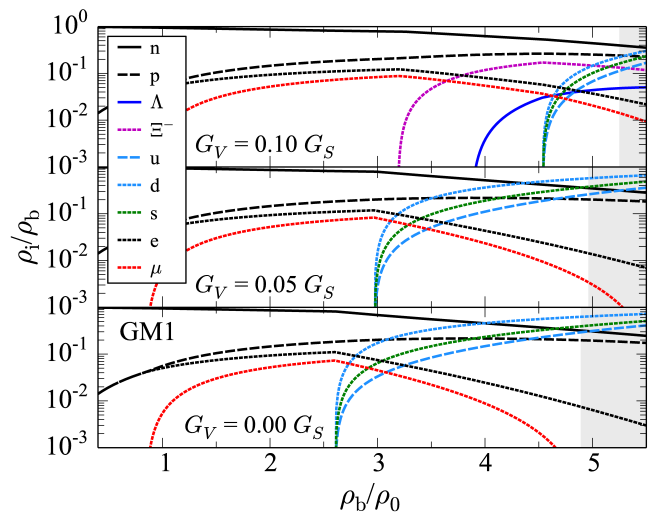}
  \caption{(Color online) Same as Figure \ref{fig:nl3_numberdensity}
  but for the GM1 parameterization.}
  \label{fig:gm1_numberdensity} 
\end{figure}

Figures \ref{fig:nl3_numberdensity} and \ref{fig:gm1_numberdensity} show
the relative particle densities for the NL3 and GM1 parameterizations and
three different values of the quark vector coupling constant. Hyperonization
does not occur at all in the NL3 parameterization, as it is preceded by
the low density onset of the quark-hadron phase transition at $2-3$
times nuclear density. The same is true of the GM1 parameterization except
in the case
that $G_V = 0.10\,G_S$. Here the onset of the quark-hadron phase transition
occurs at a much higher density due to the presence of the $\Xi^-$
and $\Lambda$ hyperons which soften the equation of state considerably, an
effect that can be seen in the right panel of Figure \ref{fig:eos}.
The low density onset of the quark-hadron phase transition is due in part
to the choice of meson-hyperon coupling constants, which have been
shown to postpone the onset of hyperonization,
stiffening the low density equation of state \cite{Miyatsu2013}. 
Figure \ref{fig:tov} shows that neutron stars within about $0.1-0.2$
$M_{\odot}$ of their maximum mass contain a quark-hadron mixed
phase in their core, with most possessing a maximum quark fraction
of around 30\% (see Table \ref{table:maxmass}).
\begin{table*}
    \begin{center}
    \caption{Properties of the maximum mass neutron star for the given
	    parameterizations.}
    \label{table:maxmass}
    \begin{tabular}{lcccccccc}
    \hline\noalign{\smallskip}
    &  & $~~$GM1$~~$ & &$~$&& $~~$NL3$~~$ & \\ \cline{2-4} \cline{6-8}
    \noalign{\smallskip}
    $G_V/G_S$ & 0 & 0.05 & 0.10 && 0 & 0.05 & 0.10 \\ 
    \hline\noalign{\smallskip}
    $M/M_{\odot}$ & 1.89 & 2.05 & 2.20 && 2.04 & 2.24 & 2.43 \\
    $\chi_{\mathrm{max}}$ & 0.32 & 0.31 & 0.15 && 0.31 & 0.30 & 0.32 \\
    $\rho_b$ [1/fm$^3$] & 0.75 & 0.76 & 0.80 && 0.61 & 0.61 & 0.62 \\
    $\epsilon$ [MeV/fm$^3$] & 851 & 883 & 969 && 687 & 696 & 726 \\
   \noalign{\smallskip}\hline
   \end{tabular}
   \end{center}
\end{table*}

\begin{figure*}[htb]
  \centering
  \includegraphics[width=0.80\textwidth]{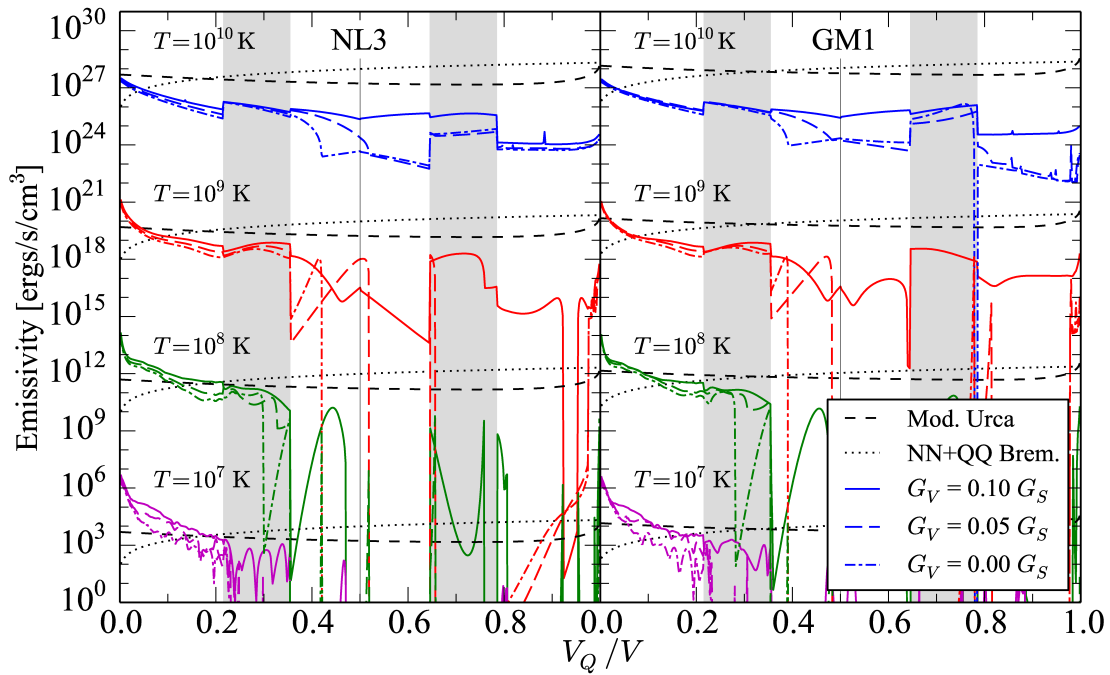}
  \caption{(Color online) Comparison of the neutrino emissivity due to
    electron-lattice interactions in the quark-hadron mixed phase to
    modified Urca and bremsstrahlung (NN+QQ) processes in the mixed
    phase for different parameterizations.  Shading represents
    geometric structures as shown in Figure \ref{fig:radii}.}
  \label{fig:emissivity}
\end{figure*}

\begin{figure*}[htb]
  \centering
  \includegraphics[width=0.80\linewidth]{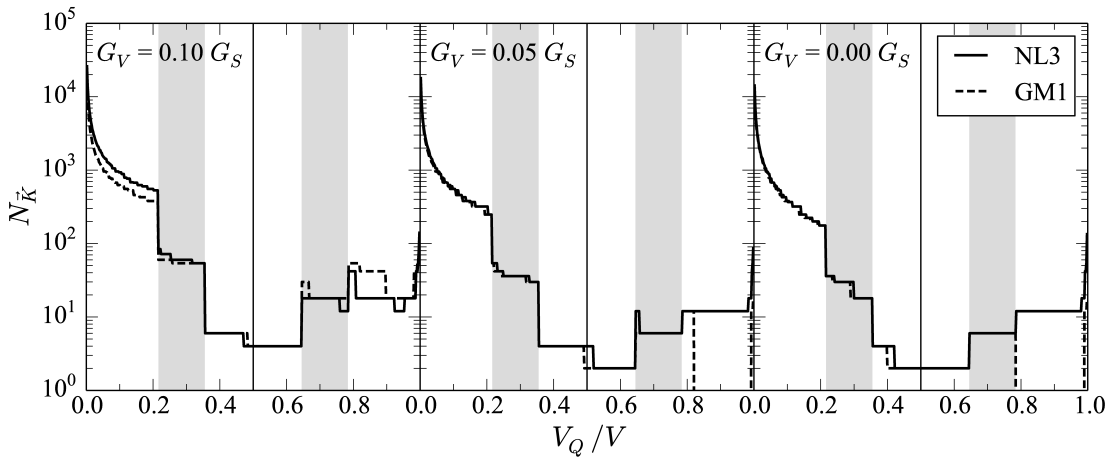}
  \caption{(Color online) The number of scattering vectors that
    satisfy the condition $K < 2k_e$ as a function of the quark
    fraction in the mixed phase for the equations of state of this
    work.}
  \label{fig:latticevectors}
\end{figure*}

Figure \ref{fig:emissivity} shows the neutrino emissivity that is due
to the crystalline structure of the quark-hadron mixed phase for all
parameterizations and temperatures between $10^7 - 10^{10}$ K,
as well as the modified Urca and bremsstrahlung (NN+QQ) emissivities
for comparison.
\begin{figure}[htb]
  \centering
  \includegraphics[width=0.95\linewidth]{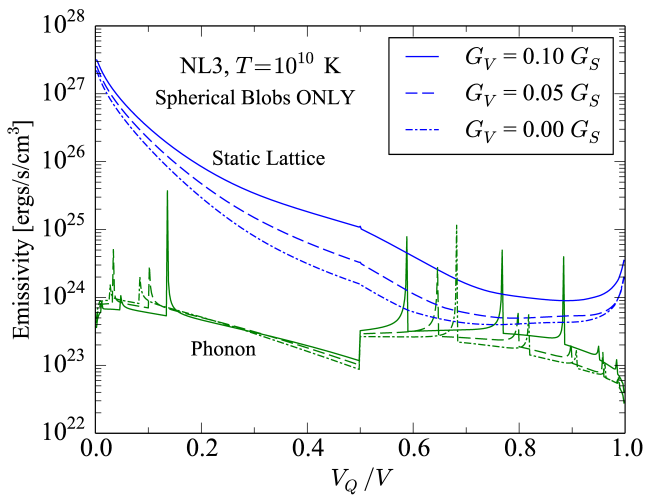}
  \caption{(Color online) Comparison of the static lattice and phonon
    contributions to the neutrino emissivity for the NL3 parameter set
    at $T > T_{\mathrm{Umklapp}}$. The comparison is performed for
    only the spherical blob geometry, as the phonon contribution is
    not determined for other geometries.}
  \label{fig:static_vs_phonon}
\end{figure}
Electron-phonon interactions contribute to the neutrino emissivity
when the mixed phase consists of spherical blobs
($\chi < 0.21$ and $\chi > 0.79$) and only at
$T > T_{\mathrm{Umklapp}}$ (Figure \ref{fig:umklapp}). Figure
\ref{fig:static_vs_phonon} shows that the static-lattice contribution
to the emissivity dominates the phonon contribution rendering it
negligible, particularly at quark fractions relevant to the neutron
stars of this work ($\chi \lesssim 0.3$).
Equations (\ref{eq:lsl}) and (\ref{eq:j}) indicate that the 
static-lattice contribution is calculated as a sum over scattering vectors
$N_{\boldsymbol{K}}$ that satisfy $K < 2k_e$. At the onset of the
mixed phase the electron Fermi momentum $k_e$ is at a maximum, which is
particularly large in magnitude due to the lack of hyperons that would
typically aid in the charge neutralization process. However, as the
quark-hadron phase transition
proceeds the negatively charged down and strange quarks take over the process of charge
neutralization, resulting in a rapidly decreasing electron number density
($k_e = (3\pi^2n_e)^{\frac{1}{3}}$). This and the exponentially decreasing
size of the Wigner Seitz cell in the spherical blob phase
($\chi < 0.21$) lead to the steep decline in
$N_{\boldsymbol{K}}$ (Figure \ref{fig:latticevectors}), which accounts for
the rapid decrease of the neutrino emissivity in the mixed phase.

The geometrical structure of the quark-hadron mixed phase terminates
with the rod phase at $\chi \sim 0.3$ for nearly all the chosen
parameterizations. Up to this point the neutrino emissivity due
to the structure of the mixed phase is either larger or comparable
to the modified Urca and bremsstrahlung (NN+QQ) emissivities for
$T \lesssim 10^8$ K.
The emissivities in the NL3 and GM1 parameterizations are comparable,
though the effect of the mixed phase structure appears more substantial
for NL3 due to lower modified Urca and bremsstrahlung (NN+QQ) emissivities.
The emissivity at very low quark fraction ($\chi \lesssim 0.05$)
may be overestimated due to the finite blob radius at $\chi = 0$ that
results from the fact that $f_d(0) = 0.4$. Finally, beyond the
rod-slab structure transition at $\chi \approx 0.35$ the electron-lattice
contribution to the overall neutrino emissivity is negligible,
though this is beyond the extent of the mixed phase of the neutron stars
in this work.


\section{\label{sec:conclusions}Summary and Conclusions}

Exploring the properties of compressed baryonic matter, or, more
generally, strongly interacting matter at high densities and/or
temperatures, has become a forefront area of modern physics
\cite{braun_munzinger09:a}.  Experimentally, such matter is being
created in relativistic particle colliders such as the Relativistic
Heavy Ion Collider RHIC at Brookhaven and the Large Hadron Collider
(LHC) at Cern, and great advances in our understanding of such matter
are expected from the next generation of collision experiments at FAIR
(Facility for Antiproton and Ion Research at GSI) and NICA
(Nucloton-based Ion Collider fAcility at JINR)
\cite{CBMbook11:a,NICA}.

Complementary to these experiments, astrophysics provides a natural
laboratory in which to explore the physics of compressed baryonic
matter too (see \cite{becker09:a,iau12:a,emmi14:a} and references
therein).  The Hubble Space Telescope and X-ray satellites such as
Chandra and XMM-Newton in particular have proven especially valuable.
New astrophysical instruments such as the Five hundred meter Aperture
Spherical Telescope (FAST), the square kilometer Array (skA), Fermi
Gamma-ray Space Telescope (formerly GLAST), Astrosat, ATHENA (Advanced
Telescope for High ENergy Astrophysics), and the Neutron Star Interior
Composition Explorer (NICER) promise the discovery of tens of
thousands of new neutron stars. Of particular interest will be the
proposed NICER mission, which is dedicated to the study of the
extraordinary gravitational, electromagnetic, and nuclear-physics
environments embodied by neutron stars. NICER will explore the exotic
states of matter in the core regions of neutron stars, confronting
nuclear theory with unique observational constraints.

With that in mind, we focus in this paper on quark deconfinement in
the cores of neutron stars.  The neutron star equation of state for
cold catalyzed matter ($T \lesssim 1$ MeV) has been determined using
the relativistic mean field (RMF) approximation to model the hadronic
phase and the nonlocal three-flavor Nambu-Jona-Lasinio model (n3NJL)
for the quark phase.  The mass-radius results indicate that a neutron
star containing quark matter in the core can account for the high mass
of the recently discovered pulsars PSR J3048+0432 and PSR J1614-2230,
and that a maximum mass neutron star can be expected to contain
approximately 30\% quark matter at the center. If the surface
tension between hadronic and quark matter is low as suggested in the
recent literature, a phase transition that results in a mixed phase
will occur in the core of a neutron star. The relaxed condition of
global charge neutrality will lead to charge segregation in the mixed
phase resulting in the formation of a crystalline lattice of quark
matter immersed in a hadronic matter background.  Expanding on Na {\it
  et al.}\ \cite{Na}, we considered the presence of two additional
geometrical structures in the mixed phase in addition to spherical
blobs: rods, and slabs (Figure \ref{fig:shapes}).

Using the formalism developed for analogous neutrino-pair
bremsstrahlung processes in the neutron star crust we have estimated
the neutrino emissivity due to electron-lattice interactions in the
quark-hadron mixed phase. The emissivity is highly dependent on the
electron number density, which has been shown to decrease considerably
in the presence of negatively charged hyperons and quarks (Figures
\ref{fig:nl3_numberdensity} and \ref{fig:gm1_numberdensity}).
However, we have shown that at temperatures
between $10^7$ K and $10^9$ K and quark fractions less than around
30\% the neutrino emissivity due to
electron-lattice interactions is significant when compared to the
standard baryon and quark modified Urca and bremsstrahlung (NN+QQ)
processes (Figure \ref{fig:emissivity}).  Further, we have also shown
that the emissivity due to electron-phonon interactions is
insignificant compared to contributions from Bragg diffraction at
temperatures above which Umklapp processes are frozen out (Figure
\ref{fig:static_vs_phonon}).

Before we can determine the effect the presence of quark matter and the
crystalline structure of the quark-hadron mixed phase has on the
thermal evolution of a neutron star the following steps need to be taken.
First, RMF should be replaced with a model 
for hadronic matter that softens the equation of state and produces
results for neutron star radii that are more compatible with
observations and recent statistical studies (see for example
\cite{steiner2010,lattimer2012}). To this end, we are currently
working on combining an RMF model that accounts for density dependence in
the values of the meson-baryon coupling constants with the three-flavor n3NJL model
(see for example 
\cite{Fuchs1995,Typel1999,Hofmann2001,Ryu2011,Colucci2013,vanDalen2014,Oertel2015,Benic2015}).
Next, the thermal conductivity and specific heat
should be calculated for the quark-hadron mixed phase as outlined in Na {\it et
al.}\ \cite{Na} using the updated equation of state for quark matter
(n3NJL) and accounting for additional rare phase geometries (rods, slabs).
Finally, these results would be incorporated into a neutron star cooling
simulation capable of properly accounting for the complexity of
the crystalline quark-hadron mixed phase.


\section*{\label{sec:acknowledgements}Acknowledgments}

This work is supported through the National Science Foundation under
grants PHY-1411708 and DUE-1259951.  Additional computing resources
are provided by the Computational Science Research Center and the
Department of Physics at San Diego State University.  GAC and MGO
acknowledge financial support by CONICET and UNLP (Project
identification code 11/G119), Argentina.


\end{document}